\pdfoutput=1

\documentclass[11pt]{article}

\usepackage[]{acl}

\usepackage{times}
\usepackage{latexsym}

\usepackage[T1]{fontenc}

\usepackage[utf8]{inputenc}

\usepackage{microtype}

\usepackage{inconsolata}

\usepackage{amsmath}
\usepackage{amssymb}
\usepackage{amsfonts}       
\usepackage{multirow}
\usepackage{makecell}
\usepackage{subcaption}
\usepackage{booktabs}       
\usepackage{nicefrac}       
\usepackage{graphicx}
\usepackage{adjustbox}
\usepackage{enumitem}
\usepackage[multiple]{footmisc}
\usepackage{hyperref}
\usepackage[ruled,linesnumbered,vlined]{algorithm2e}

\usepackage{authblk}
\makeatletter
\renewcommand\AB@affilsepx{\quad\protect\Affilfont} 
\makeatother
\newcommand\blfootnote[1]{%
  \begingroup
  \renewcommand\thefootnote{}\footnote{#1}%
  \addtocounter{footnote}{-1}%
  \endgroup
}
\title{Aligning Large Language Models with Recommendation Knowledge}
\author[1*]{Yuwei Cao}
\author[2]{Nikhil Mehta}
\author[2]{Xinyang Yi}
\author[3]{Raghunandan Keshavan}
\author[3]{\authorcr Lukasz Heldt}
\author[2]{Lichan Hong}
\author[2]{Ed H. Chi}
\author[4]{Maheswaran Sathiamoorthy}
\affil[1]{University of Illinois Chicago}
\affil[2]{Google DeepMind}
\affil[3]{Google \protect\\}
\affil[1]{\href{mailto:ycao43@uic.edu}{\texttt{ycao43@uic.edu}}}
\affil[2]{\{\href{mailto:nikhilmehta@google.com}{\texttt{nikhilmehta}}, \href{mailto:xinyang@google.com}{\texttt{xinyang}}\}\texttt{@google.com} \protect\\}
\affil[4]{\href{mailto:mahesh@smahesh.com}{\texttt{mahesh@smahesh.com}}}

\begin{document}
\maketitle
\blfootnote{\textsuperscript{*}Work done when interning at Google.}
\begin{abstract}

Large language models (LLMs) have recently been used as backbones for recommender systems. However, their performance often lags behind conventional methods in standard tasks like retrieval. We attribute this to a mismatch between LLMs' knowledge and the knowledge crucial for effective recommendations. While LLMs excel at natural language reasoning, they cannot model complex user-item interactions inherent in recommendation tasks. We propose bridging the knowledge gap and equipping LLMs with recommendation-specific knowledge to address this. 
Operations such as Masked Item Modeling (MIM) and Bayesian Personalized Ranking (BPR) have found success in conventional recommender systems. Inspired by this, we simulate these operations through natural language to generate auxiliary-task data samples that encode item correlations and user preferences.
Fine-tuning LLMs on such auxiliary-task data samples and incorporating more informative recommendation-task data samples facilitates the injection of recommendation-specific knowledge into LLMs. Extensive experiments across retrieval, ranking, and rating prediction tasks on LLMs such as FLAN-T5-Base and FLAN-T5-XL show the effectiveness of our technique in domains such as Amazon Toys \& Games, Beauty, and Sports \& Outdoors. Notably, our method outperforms conventional and LLM-based baselines, including the current SOTA, by significant margins in retrieval, showcasing its potential for enhancing recommendation quality.


\end{abstract}
\section{Introduction}

\begin{figure*}[t]
\centering
\includegraphics[width = 16cm]{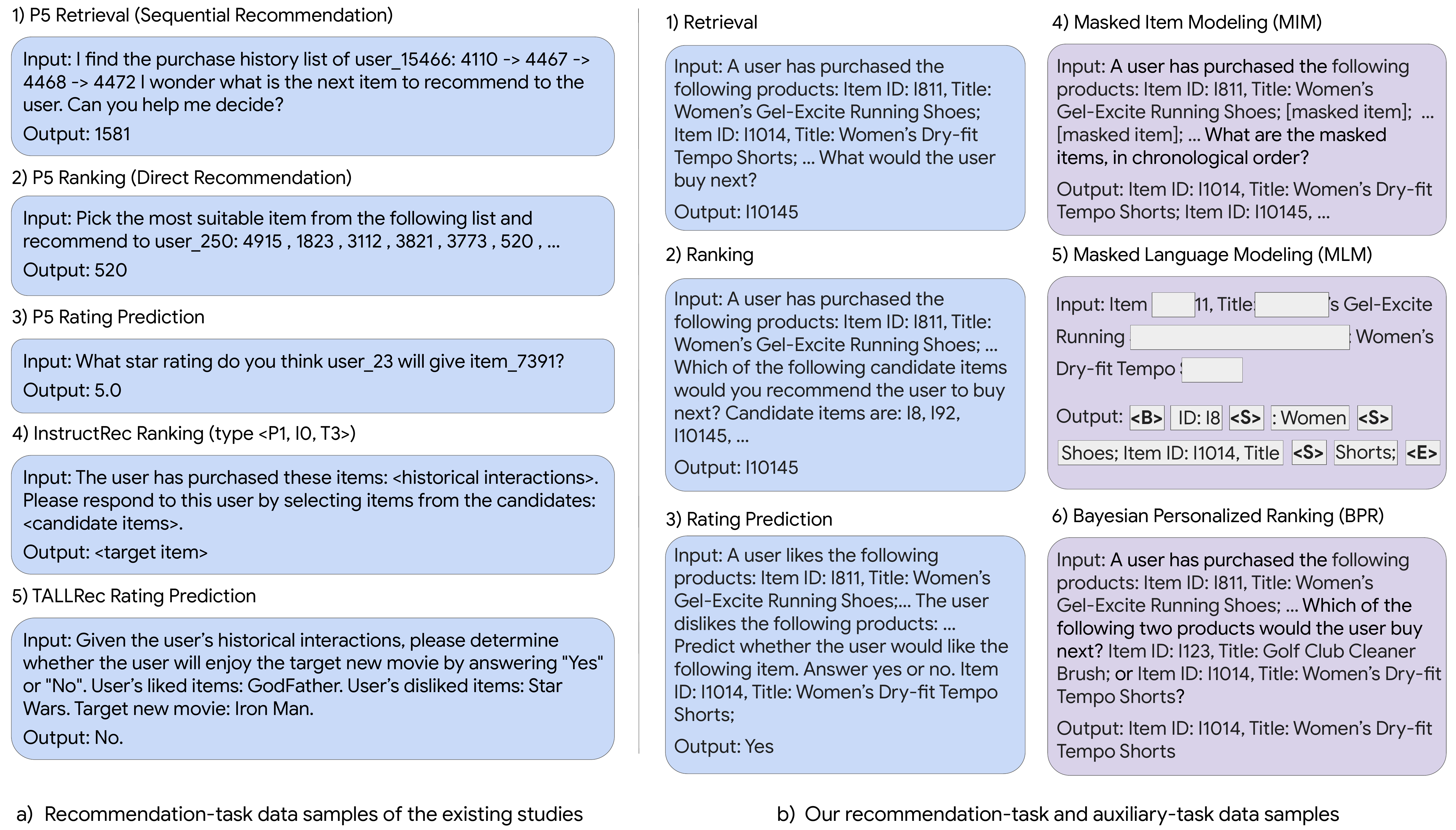}
\caption{\label{fig:prompts} Data samples adopted by the existing studies and this work. (a) shows the recommendation-task data samples of the existing studies. Specifically, (a1)-(a3) demonstrate the retrieval, ranking, and  rating prediction data samples of P5 \cite{geng2022recommendation}; (a4) shows a ranking (type <P1, I0, T3>) data sample of InstructRec \cite{zhang2023recommendation}; (a5) is a rating prediction data sample of TALLRec \cite{bao2023tallrec}. (b) shows our recommendation-task (blue boxes) and auxiliary-task (purple boxes) data samples (we present more samples in Appendix \ref{sec:appendix_samples}).}
\label{fig:overview}
\vspace{-0.5cm}
\end{figure*}

Large language models (LLMs) exhibit strong generalization abilities through zero-shot learning, in-context learning \cite{brown2020language}, fine-tuning, and instruction tuning 
\cite{wei2021finetuned}. Encouraged by this, recent studies explore the use of LLMs as backbones in recommendation \cite{kang2023llms, geng2022recommendation, zhang2023recommendation, bao2023tallrec}. Despite their great potential, LLMs are inferior to supervised recommenders \cite{he2017neural, rendle2012bpr} in recommendation tasks such as rating-prediction under zero-shot and few-shot in-context learning settings \cite{kang2023llms}. We hypothesize that this stems from a gap between LLMs' knowledge and recommendation knowledge: LLMs are proficient at natural language reasoning, while recommendation involves modeling complex user-item interactions.
In this work, we propose to mitigate this gap by fine-tuning LLMs with data samples that encode recommendation knowledge.

Recent works \cite{geng2022recommendation, zhang2023recommendation, bao2023tallrec} show that certain recommendation knowledge can be introduced into LLMs through instruction tuning.
As shown in Figure \ref{fig:overview}(a), their training data samples, which we refer to as \textit{recommendation-task data samples}, primarily help LLMs understand the recommendation tasks by providing instructions on what to do (\textit{e.g.}, ``Pick an item for the user from the following candidates.''). In terms of modeling the target recommendation domain, however, they present raw user and item features for personalization (\textit{e.g.}, the user's ID or the IDs of the items they recently interacted with), which are insufficient for LLMs to fully comprehend the target domain. 

Considering the aforementioned limitations of using LLMs as recommenders, we propose a novel approach to generate additional fine-tuning data samples for LLMs that effectively encode recommendation knowledge, particularly focusing on item correlations within the target domain. We refer to these generated data samples as \textit{auxiliary-task data samples}, as they are used as auxiliary tasks in addition to the recommendations tasks. 
While developing the auxiliary tasks, our key inspiration comes from the classical operations that are typically used to train conventional recommender systems, namely, masked item modeling (MIM)~\cite{sun2019bert4rec} and Bayesian Personalized Ranking (BPR)~\cite{rendle2012bpr}. Our key innovation lies in converting the MIM and BPR tasks into natural language tasks that can be used to train the LLMs. We also incorporate the masked language modeling (MLM)~\cite{devlin2018bert} task for the user's past interactions to supplement the MIM task with fine-grained item correlations.
\noindent Our contributions can be summarized as follows:
\begin{itemize}[leftmargin=4pt,noitemsep,nolistsep]
    \setlength{\itemsep}{4pt}
    \item We propose a novel method to align LLMs with new recommendation domains, \textit{i.e.}, supplementing the fine-tuning of the LLMs with \textit{auxiliary-task data samples} that mimic the classical operations in training conventional recommender systems with natural language prompts.
    \item We propose \textit{recommendation-task data samples} that are more informative as compared to the existing work \cite{geng2022recommendation}. Specifically, we reduce the complexity of the input/output spaces by eliminating the user IDs. We further enhance the user sequences by providing item titles.

    \item We fine-tune the open-source 3B FLAN-T5-XL and 223M FLAN-T5-Base with our proposed recommendation-task and auxiliary-task data samples in a simple multi-task learning framework. Experiments on various recommendation tasks, \textit{i.e.}, retrieval, ranking, and rating-prediction, across three target domains, \textit{i.e.}, Amazon Toys \& Games, Beauty, and Sports \& Outdoors, show the effectiveness of our proposed method and its components. For retrieval, our model outperforms both conventional and LLM-based baselines, including the current SOTA, by large margins.
\end{itemize}

\section{Related Work}
\textbf{Recommender Systems.} 
Recommender systems help users in discovering items of interest. As a practical approach, Collaborative Filtering (CF) \cite{mao2021simplex} explores historical user-item interactions, assuming that users with similar behaviors have similar preferences for items. Among various CF methods, Matrix Factorization (MF) methods \cite{rendle2012bpr, mao2021simplex} project users and items into a shared vector space and estimate a user's preference for an item through the inner product of their vectors and are widely adopted. Context-aware approaches \cite{cheng2016wide} further include additional information, such as user and contextual features, to improve recommendation quality. However, CF fails to capture the sequential patterns in users' behaviors, which leads to the rise of sequential recommendations. Sequential recommenders based on Convolutional Neural Networks (CNNs) \cite{tang2018personalized}, Gated Recurrent Units (GRUs) \cite{hidasi2015session}, and self-attention \cite{sun2019bert4rec, zhang2019feature, kang2018self, zhou2020s3, rajput2023recommender} have become prevalent in the era of deep learning. 
Notably, leveraging a T5-like backbone, \citealt{rajput2023recommender} formalize recommendation as generative retrieval, \textit{i.e.}, autoregressively decode the identifiers of the target items, and achieve the current SOTA. While structurally resembling LLMs, it lacks their pre-training knowledge and the accompanying natural language reasoning potential. 
Our proposed approach adopts self-attention for sequential recommendation, specifically harnessing LLMs as backbones. We compare against various baselines from all the classes discussed above.
\vspace{0.5\baselineskip}

\noindent\textbf{LLMs for Recommendation.}
LLMs have recently been explored for recommendation tasks due to their ability to understand, generate, and reason with natural language. Several studies focus on incorporating LLMs' natural language capabilities into existing recommendation techniques. \textit{E.g.}, \citealt{hou2022towards} and \citealt{cao2023multi} encode item contents (title, description, etc.) with BERT \cite{devlin2018bert}, which enables learning semantically informed embeddings even for zero-shot items. Moreover, pre-trained LLM backbones have also been used for recommendation through zero-shot learning \cite{kang2023llms}, in-context learning \cite{kang2023llms}, fine tuning \cite{cui2022m6, kang2023llms}, and instruction tuning \cite{geng2022recommendation, zhang2023recommendation, bao2023tallrec}. Besides helping classic recommendation tasks, LLMs also enable novel recommendation use cases. \citealt{geng2022recommendation} leverage LLMs to explain the recommendation results.
\citealt{gao2023chat, wang2023zero} utilize GPT-3 \cite{brown2020language} for conversational recommendation. \citealt{christakopoulou2023large} extract persistent user interests with LLMs for deeper user understanding. \citealt{carranza2023privacy} generate private synthetic representations of the original data with LLMs for privacy-preserving recommendation.
\vspace{0.5\baselineskip}

\noindent\textbf{Recommendation as Instruction-following.} The success of instruction tuning, \textit{i.e.}, fine-tune on data described via instructions \cite{mishra2021cross, wei2021finetuned}, has inspired attempts that instruction-tune LLM backbones for recommendation tasks. \citealt{geng2022recommendation} formalize various recommendation tasks as natural language instructions and fine-tune a unified recommender with T5 \cite{raffel2020exploring} backbone. 
\citealt{zhang2023recommendation} further supplement the tuning data with user preferences/intentions deduced by GPT-3.5 \footnote{\url{https://platform.openai.com/docs/models/overview}} to accommodate instructions of free forms.
\citealt{bao2023tallrec} explore instruction tuning LLMs with limited data.

In contrast to the existing studies, our work focuses on introducing new recommendation knowledge into LLMs, which we believe is the key for improving recommenders with LLM backbones. We create auxiliary tasks that improve the recommendation tasks, including retrieval, ranking, and rating prediction. 
Our proposed recommendation-task and auxiliary-task data samples include raw user purchase sequences in addition to natural language instructions. These data samples supplement each other in encoding the target recommendation domain knowledge. We experiment under restricted settings. Compared to the previous studies \cite{zhang2023recommendation}, we consider larger candidate pools (\textit{e.g.}, our retrieval and ranking experiments consider the entire dataset and 99 hard negatives, respectively). Unlike \citealt{bao2023tallrec}, we fully train all models to maximize their performances. 

\section{Methodology}
We propose designing data samples that encode recommendation knowledge to align LLMs with the target recommendation domain. Sections \ref{sec:AT_generation} and \ref{sec:RT_generation} discuss our auxiliary-task and recommendation-task data, respectively. Section \ref{sec:instruct_tuning_framework} introduces a simple multi-task learning framework that we use to fine-tune LLMs.


\subsection{Auxiliary-task Data Generation}
\label{sec:AT_generation}
Conventional recommenders acquire recommendation knowledge via classic operations such as masked item modeling \cite{sun2019bert4rec} and BPR loss reduction \cite{rendle2012bpr}. We mimic these operations with natural language prompts. In addition, we sample sub-sequences of the raw user purchase sequences. The resulting data, which we refer to as auxiliary-task data samples, encode item correlations contained in users' preferences \footnote{As a side note, we also explored encoding item correlations contained in item contents (categories, descriptions, etc.). Observing no noticeable performance increase, we present our approach and results in Appendix \ref{sec:appendix_IE}}.
\subsubsection{Masked Item Modeling (MIM)}
\label{sec:MIM}
Conventional sequential recommenders \cite{sun2019bert4rec} learn item correlations from users' interaction sequences. Specifically, they predict randomly masked items in the sequences by jointly conditioning on the unmasked items. We mimic this process, which we refer to as masked item modeling (MIM), with natural language prompts.


MIM applies a Cloze objective \cite{sun2019bert4rec}. At each training step, random items in the input user sequence are replaced with a special token "[mask]", and the model learns to recover the masked items based on its surrounding context. An example of the masking process:
\begin{equation}
\begin{split}
\text{\textbf{Input: }} & \text{[}i_1, i_2, i_3, i_4, i_5\text{]} \xrightarrow[]{\text{random masking}} \\ 
& \text{[}i_1, \text{[mask]}_1, i_3, \text{[mask]}_2, i_5\text{]} \\
\text{\textbf{Label: }} & \text{[mask]}_1 = i_2,\;\text{[mask]}_2 = i_4
\end{split}
\end{equation}

The MIM loss is computed as follows in conventional sequential recommenders:
\begin{equation}
\label{eq:mim_loss}
\mathcal{L}_{\mathrm{MIM}} = \frac{1}{|\mathcal{S}_u^m|}\sum_{i_m \in \mathcal{S}_u^m}-\text{log}P(i_m|\mathcal{S}_u^{'}),
\end{equation}
where $\mathcal{S}_u^{'}$ is the masked version of user sequence $\mathcal{S}_u$, $\mathcal{S}_u^m$ stands for the masked items in $\mathcal{S}_u$. $P(\cdot)$, the probability of observing $i_m$ given $\mathcal{S}_u^{'}$, is calculated from deep bidirectional self-attention \cite{devlin2018bert}.

Our natural language imitation of MIM loss (Equation \ref{eq:mim_loss}) is described in Figure~\ref{fig:overview}(b4). Given purchase sequence: $\text{[}i_1, i_2, i_3, i_4, i_5\text{]}$, we generate prompts, \textit{e.g.}, Input: ``A user has purchased the following products: Item ID: $[\textrm{ID}]_{i_1}$, Title: $[\textrm{Title}]_{i_1}$; [masked item]; Item ID: $[\textrm{ID}]_{i_3}$, Title: $[\textrm{Title}]_{i_3}$; [masked item]; Item ID: $[\textrm{ID}]_{i_5}$, Title: $[\textrm{Title}]_{i_5}$. What are the masked items, in chronological order?'', and Output: ``Item ID: $[\textrm{ID}]_{i_2}$, Title: $[\textrm{Title}]_{i_2}$; Item ID: $[\textrm{ID}]_{i_4}$, Title: $[\textrm{Title}]_{i_4}$;''. 
To accommodate long sequences, we introduce a sliding window $w$ and each prompt considers one sub-sequence: $\text{[}i_k, i_{k+1}..., i_{k+w-1}\text{]}$, where $1 \le k \le \max\bigl($1,$(L$-$w$+1$)\bigr)$ and $L$ is the total length of the user sequence. The resulting MIM data samples encodes the correlations between the masked items and the rest of the sequences.

\subsubsection{Masked Language Modeling (MLM)}
In addition to MIM that considers a single item for each mask, we also mask out and recover a consecutive span of \textit{tokens} to encode fine-grained item correlations contained in the users' purchase sequences. This process resembles masked language modeling (MLM) \cite{devlin2018bert}.

As shown in Figure~\ref{fig:overview}(b5), given a user sequence, we sample a sub-sequence by randomly deciding a starting item and a sub-sequence length $L_s$, where 2 $\le L_s \le w$ and $w$ is the sliding window for accommodating long sequences. These sub-sequences, referred to as MLM data samples, supplement the MIM data samples: through span corruption \cite{raffel2020exploring}, \textit{i.e.}, masking and recovering consecutive spans of tokens, LLMs learn to model more fine-grained correlations across multiple continuous items from the MLM data samples.


\subsubsection{Bayesian Personalized Ranking (BPR)}
Besides correlating similar items, we explore contrasting dissimilar items. BPR loss \cite{rendle2012bpr} is adopted by conventional recommenders \cite{rendle2014improving, koren2009matrix, cheng2016wide} for personalized ranking, \textit{i.e.}, learning users' preferences for some items over the others.
Inspired by this, we imitate BPR loss reduction with natural language prompts for training LLMs.


The objective of BPR loss reduction in conventional recommenders is:
\begin{equation}
\mathcal{L}_{\mathrm{BPR}} = \mathop{\mathbb{E}}_{(u,i^+)\sim p_{\mathrm{pos}}}-\log\sigma(s(u,i^+)-s(u,i^-)),
\end{equation}
where $(u,i^+)$ is a pair of a user $u$ and an item $i^+$ sampled from the distribution of positive pairs $p_{\text{pos}}$, \textit{i.e.}, $u$ interacted with $i^+$. $i^-$ is a randomly sampled negative item that $u$ has not interacted with. The similarity between $u$ and $i^+$, denoted by $s(u,i^+)$, is calculated by taking the dot product of their representations. $\sigma(\cdot)$ is the Sigmoid function.

Figure~\ref{fig:overview}(b6) shows our natural language imitation. We elicit user preferences by generating prompts with binary choices that contrast a positive item and a negative item. Each prompt takes the form of a binary decision, \textit{e.g.}, Input: ``A user has purchased ... Which of the following two products would the user buy next? Item ID: $[\textrm{ID}]_{i^-}$, Title: $[\textrm{Title}]_{i^-}$; Item ID: $[\textrm{ID}]_{i^+}$, Title: $[\textrm{Title}]_{i^+}$.'', and Output: ``Item ID: $[\textrm{ID}]_{i^+}$, Title: $[\textrm{Title}]_{i^+}$''. Following Section~\ref{sec:MIM}, we adopt a sliding window $w$ to accommodate long user sequences and the positive item is always the one next to the sliding window. These BPR data samples encode dissimilarities between the purchased items and the rest of the items in the dataset.

\subsection{Recommendation-task Data Generation}
\label{sec:RT_generation}
As shown in Figure~\ref{fig:prompts}(a), the existing recommenders with LLM backbones adopt prompts that primarily convey the recommendation tasks by providing directions on how to perform them. Such information is essential, yet insufficient for representing the target recommendation domain. 

We propose prompts that help LLMs comprehend the target recommendation domain in addition to the recommendation tasks. Specifically, we reduce the complexity of the input/output spaces. In contrast to \citealt{geng2022recommendation}, we eliminate user IDs and represent the users by their historical purchases. Consequently, we relieve LLMs from memorizing a substantial volume of user IDs (\textit{e.g.}, Amazon Sports \& Outdoors has 35,598 users). 
Moreover, compared to \citealt{geng2022recommendation} that represent user sequences solely by item IDs, we include both the IDs and the titles of the items, which makes it easier for LLMs to recognize the items. Notably, ranking candidates and items in the output are represented solely by their IDs to reduce the length of the prompts and maintain a smaller output space.
Figures~\ref{fig:prompts}(b1)-(b3) show examples of our retrieval, ranking, and rating prediction recommendation-task data samples. The raw item IDs (\textit{e.g.}, `0000031852') are mapped into shorter ones (\textit{e.g.}, `I123') \footnote{We adopt random mapping, \textit{i.e.}, similar-looking IDs may not imply any connection or semantic similarity. We acknowledge that using semantic-rich IDs \cite{rajput2023recommender} could enhance performance and leave the exploration to the future.} to reduce input/output space complexity. To fully present the users' historical purchases to LLMs, we adopt a sliding window $w$ similar to Section~\ref{sec:MIM}.

\subsection{Fine-tuning and Evaluation Framework}
\label{sec:instruct_tuning_framework}

\begin{figure}[t]
\centering
\includegraphics[width = 7.3cm]{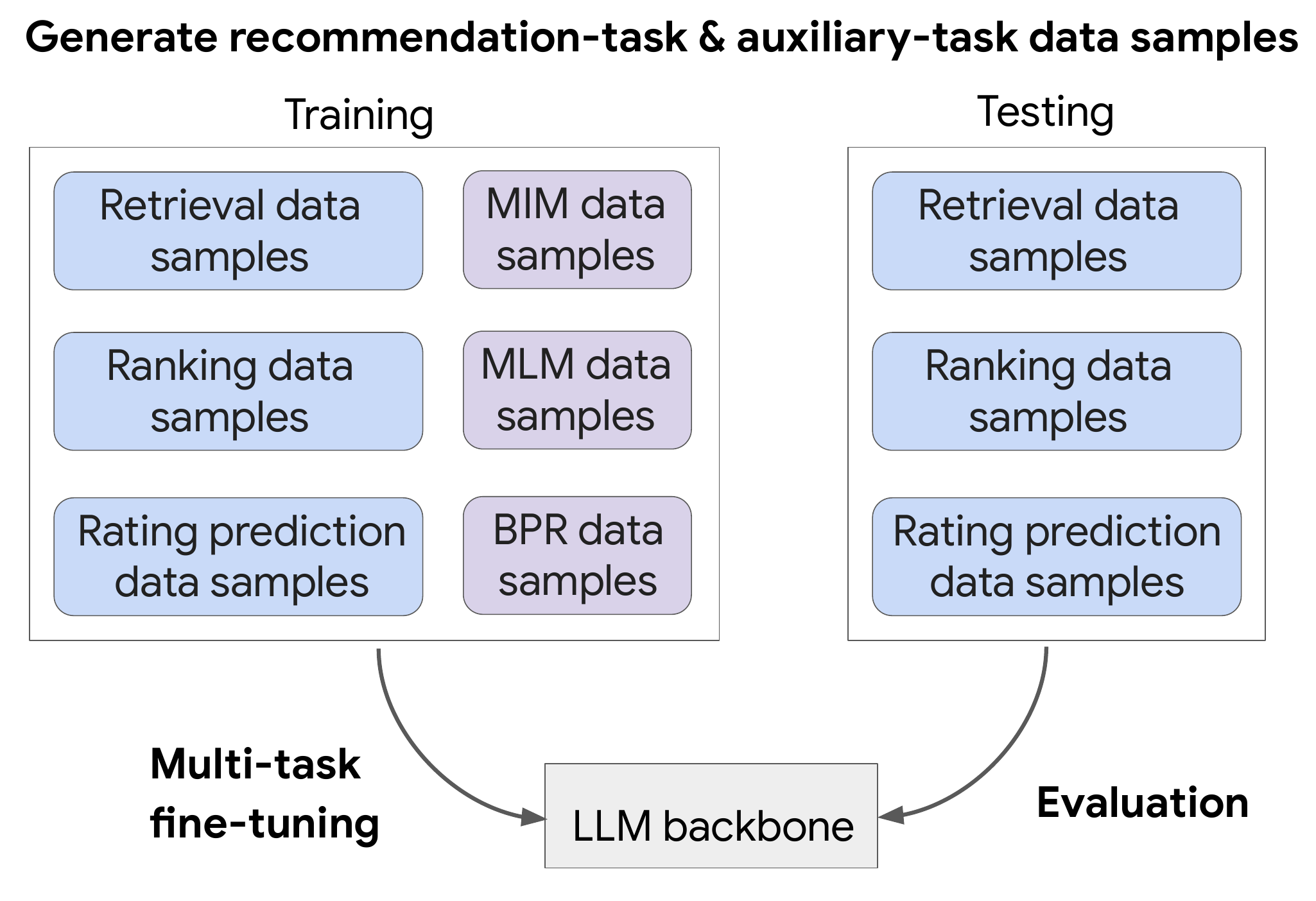}
\caption{\label{fig:framework} Fine-tuning and evaluation framework.}
\vspace{-0.5cm}
\end{figure}

As shown in Figure \ref{fig:framework}, we adopt a simple framework to fine-tune the LLM backbones and evaluate the resulting model. We first generate recommendation-task and auxiliary-task data samples using the training set. Next, we tune the LLM backbone with these data samples in a multi-task learning manner. Finally, we evaluate the recommendation tasks using the recommendation-task data samples generated from the test set.

\begin{table*}[t]
\centering
\begin{adjustbox}{width=\linewidth}
\begin{tabular}{lcccccccccccc}
    \toprule
    \multirow{2.5}{*}{Methods} &
    \multicolumn{4}{c}{\textbf{Toys \& Games}} & \multicolumn{4}{c}{\textbf{Beauty}} & \multicolumn{4}{c}{\textbf{Sports \& Outdoors}} \\
   \cmidrule(lr){2-5}\cmidrule(lr){6-9}\cmidrule(lr){10-13}
   & \shortstack{NDCG\\@5}  & \shortstack{NDCG\\@10} & \shortstack{HR\\@5}  & \shortstack{HR\\@10} & \shortstack{NDCG\\@5}  & \shortstack{NDCG\\@10} & \shortstack{HR\\@5}  & \shortstack{HR\\@10} & \shortstack{NDCG\\@5}  & \shortstack{NDCG\\@10} & \shortstack{HR\\@5}  & \shortstack{HR\\@10}  \\
   \cmidrule{1-13}
   Caser$^1$ & 0.0107 & 0.0141 & 0.0166 & 0.0270 & 0.0131 & 0.0176 & 0.0205 & 0.0347 & 0.0072 & 0.0097 & 0.0116 & 0.0194 \\
   HGN$^1$ & 0.0221 & 0.0277 & 0.0321 & 0.0497 & 0.0206 & 0.0266 & 0.0325 & 0.0512 & 0.0120 & 0.0159 & 0.0189 & 0.0313 \\
   GRU4Rec$^1$ & 0.0059 & 0.0084 & 0.0097 & 0.0176 & 0.0099 & 0.0137 & 0.0164 & 0.0283 & 0.0086 & 0.0110 & 0.0129 & 0.0204 \\ 
   BERT4Rec$^1$ & 0.0071 & 0.0099 & 0.0116 & 0.0203 & 0.0124 & 0.0170 & 0.0203 & 0.0347 & 0.0075 & 0.0099 & 0.0115 & 0.0191 \\ 
   FDSA$^1$ & 0.0140 & 0.0189 & 0.0228 & 0.0381 & 0.0163 & 0.0208 & 0.0267 & 0.0407 & 0.0122 & 0.0156 & 0.0182 & 0.0288 \\
   SASRec$^1$ & 0.0306 & 0.0374 & 0.0463 & 0.0675 & 0.0249 & 0.0318 & 0.0387 & 0.0605 & 0.0154 & 0.0192 & 0.0233 & 0.0350 \\
   S$^3$-Rec$^1$ & 0.0294 & 0.0376 & 0.0443 & 0.0700 & 0.0244 & 0.0327 & 0.0387 & 0.0647 & 0.0161 & 0.0204 & 0.0251 & 0.0385 \\
   TIGER$^2$ & 0.0371 & 0.0432 & 0.0521 & 0.0712 & 0.0321 & 0.0384 & 0.0454 & 0.0648 & \underline{0.0181} & \underline{0.0225} & 0.0264 & \underline{0.0400} \\
   \midrule
   P5$^2$ & 0.0050 & 0.0066 & 0.0070 & 0.0121 & 0.0107 & 0.0136 & 0.0163 & 0.0254 & 0.0041 & 0.0052 & 0.0061 & 0.0095 \\
   P5-XL & 0.0023 & 0.0031 & 0.0035 & 0.0061 & 0.0036 & 0.0050 & 0.0063 & 0.0104 & 0.0029 & 0.0035 & 0.0040 & 0.0060 \\
   FLAN-T5-Base & 0.0000 & $2\mathrm{e}{-5}$ & 0.0000 & $5\mathrm{e}{-5}$ & 0.0000 & 0.0000 & 0.0000 & 0.0000 & 0.0000 & $9\mathrm{e}{-6}$ & 0.0000 & $3\mathrm{e}{-5}$\\
   FLAN-T5-XL & $2\mathrm{e}{-5}$ & $2\mathrm{e}{-5}$ & $5\mathrm{e}{-5}$ & $5\mathrm{e}{-5}$ & 0.0000 & 0.0000 & 0.0000 & 0.0000 & 0.0000 & 0.0000 & 0.0000 & 0.0000 \\
   \midrule
   ReAT [\textbf{Ours}] & \underline{0.0390} & \textbf{0.0461} & \underline{0.0558} & \textbf{0.0776} & \textbf{0.0382} & \textbf{0.0442} & \textbf{0.0535} & \textbf{0.0722} & \textbf{0.0188} & \textbf{0.0232} & \textbf{0.0285} & \textbf{0.0422} \\
   UT [\textbf{Ours}] & 0.0166 & 0.0202 & 0.0252 & 0.0362 & 0.0188 & 0.0231 & 0.0292 & 0.0425 & 0.0079 & 0.0101 & 0.0118 & 0.0187 \\
   UT+AT [\textbf{Ours}] & \textbf{0.0392} & \underline{0.0459} & \textbf{0.0563} & \underline{0.0772} & \underline{0.0329} & \underline{0.0397} & \underline{0.0482} & \underline{0.0693} & 0.0178 & 0.0219 & \underline{0.0268} & 0.0393 \\
   \midrule
   $\Delta$ (\%) & \textcolor{blue}{+5.66} & \textcolor{blue}{+6.71} & \textcolor{blue}{+8.06} & \textcolor{blue}{+8.99} & \textcolor{blue}{+19.00} & \textcolor{blue}{+15.10} & \textcolor{blue}{+17.84} & \textcolor{blue}{+11.42} & \textcolor{blue}{+3.87} & \textcolor{blue}{+3.11} & \textcolor{blue}{+7.95} & \textcolor{blue}{+5.50} \\
   \bottomrule
   \end{tabular}
\end{adjustbox}
\caption{Retrieval results. \textsuperscript{1} marks results from \citealt{zhou2020s3}; $^2$ marks results from \citealt{rajput2023recommender}. $\Delta$ compares the best [\textbf{Ours}] with the best baseline.}
\label{table:retrieval_results}
\end{table*}

\begin{table*}[t]
\centering
\begin{adjustbox}{width=\linewidth}
\begin{tabular}{lccccccccccccccc}
    \toprule
    \multirow{2.5}{*}{Methods} &
    \multicolumn{5}{c}{\textbf{Toys \& Games}} & \multicolumn{5}{c}{\textbf{Beauty}} & \multicolumn{5}{c}{\textbf{Sports \& Outdoors}} \\
   \cmidrule(lr){2-6}\cmidrule(lr){7-11}\cmidrule(lr){12-16}
   & \shortstack{NDCG\\@5}  & \shortstack{NDCG\\@10} & \shortstack{HR\\@1} & \shortstack{HR\\@5}  & \shortstack{HR\\@10}
   & \shortstack{NDCG\\@5}  & \shortstack{NDCG\\@10} & \shortstack{HR\\@1} & \shortstack{HR\\@5}  & \shortstack{HR\\@10}
   & \shortstack{NDCG\\@5}  & \shortstack{NDCG\\@10} & \shortstack{HR\\@1} & \shortstack{HR\\@5}  & \shortstack{HR\\@10} \\
   \cmidrule{1-16}
   BPR-MF\textsuperscript{1}  & 0.0641 & 0.0940  & 0.0233 & 0.1066 & 0.2003 & 0.0857 & 0.1224 & 0.0311 & 0.1426 & 0.2573 & 0.0848 & 0.1220 & 0.0314 & 0.1404 & 0.2563\\
   BPR-MLP\textsuperscript{1} & 0.0688 & 0.0988 & 0.0252 & 0.1142 & 0.2077 & 0.0848 & 0.1215 & 0.0317 & 0.1392 & 0.2542 & 0.0927 & \underline{0.1296} & 0.0351 & 0.1520 & \underline{0.2671}\\
   SimpleX\textsuperscript{1}  & 0.1244 & 0.1469 & 0.0268 & 0.1958 & 0.2662 & \textbf{0.1441} & \underline{0.1711} & 0.0325 & \textbf{0.2247} & \underline{0.3090} & \textbf{0.1505} & \textbf{0.1800} & 0.0331 & \textbf{0.2362} & \textbf{0.3290}\\
   \midrule
   P5-XL & 0.0290 & 0.0444 & 0.0097 & 0.0494 & 0.0977 & 0.0298 & 0.0456 & 0.0110 & 0.0498 & 0.0992 & 0.0286 & 0.0436 & 0.0097 & 0.0486 & 0.0957 \\
   FLAN-T5-Base  & 0.0107 & 0.0127 & 0.0057 & 0.0156 & 0.0217 & 0.0097 & 0.0113 & 0.0052 & 0.0137 & 0.0189 & 0.0069 & 0.0082 & 0.0035 & 0.0102 & 0.0144\\
   FLAN-T5-XL & 0.0160 & 0.0312 & 0.0026 & 0.0315 & 0.0793 & 0.0152 & 0.0296 & 0.0022 & 0.0301 & 0.0753 & 0.0097 & 0.0193 & 0.0014 & 0.0192 & 0.0491\\
   \midrule
   RaAT [\textbf{Ours}] & \textbf{0.1714} & \underline{0.2034} & \textbf{0.0956} & \textbf{0.2464} & \underline{0.3453} & \underline{0.1376} & 0.1691 & \underline{0.0702} & 0.2036 & 0.3013 & 0.0933 & 0.1199 & \underline{0.0424} & 0.1448 & 0.2272 \\
   UT [\textbf{Ours}] & 0.1536 & 0.1867 & 0.0831 & 0.2233 & 0.3259 & 0.1236 & 0.1537 & 0.0609 & 0.1863 & 0.2798 & 0.0867 & 0.1137 & 0.0381 & 0.1362 & 0.2202 \\
   UT+AT [\textbf{Ours}] & \underline{0.1703} & \textbf{0.2064} & \underline{0.0938} & \underline{0.2443} & \textbf{0.3562} & \textbf{0.1441} & \textbf{0.1758} & \textbf{0.0742} & \underline{0.2126} & \textbf{0.3112} & \underline{0.0997} & 0.1281 & \textbf{0.0468} & \underline{0.1526} & 0.2404 \\
   \midrule
   $\Delta$ (\%) & \textcolor{blue}{+37.78} & \textcolor{blue}{+40.50} & \textcolor{blue}{+256.72} & \textcolor{blue}{+25.84} & \textcolor{blue}{+33.81} & 0.00 & \textcolor{blue}{+2.75} & \textcolor{blue}{+128.31} & -5.38 & \textcolor{blue}{+0.71} & -33.75 & -28.83 & \textcolor{blue}{+33.33} & -35.39 & -26.93 \\
   \bottomrule
   \end{tabular}
   \end{adjustbox}
\caption{Ranking results. \textsuperscript{1} marks results from \citealt{geng2022recommendation}. $\Delta$ compares the best [\textbf{Ours}] with the best baseline.}
\label{table:ranking_results}
\vspace{-0.3cm}
\end{table*}


\begin{table}[t]
\centering
\begin{adjustbox}{width=0.95\linewidth}
\begin{tabular}{lccc}
    \toprule
    Methods & \textbf{Toys \& Games} & \textbf{Beauty} & \textbf{Sports \& Outdoors}
    \\
    \midrule
    History & 66.59 & 64.80 & 62.78 \\
    DMF & 51.82 & 51.23 & 51.38 \\ 
    Wide\&Deep & 70.93 & 67.10 & \textbf{67.60} \\
    P5-XL & 51.04 & 50.63 & 50.36 \\ 
    FLAN-T5-Base & 57.85 & 56.04 & 55.00 \\ 
    FLAN-T5-XL & 55.23 & 53.77 & 52.01 \\
    \midrule
    RpAT [\textbf{Ours}] & \textbf{71.16} & \textbf{68.27} & \underline{65.87} \\
    UT [\textbf{Ours}] & 70.79 & 67.45 & 65.35\\
    UT+AT [\textbf{Ours}] & \underline{71.08} & \underline{67.55} & 65.18 \\
    \midrule
   $\Delta$ (\%) & \textcolor{blue}{+0.32} & \textcolor{blue}{+1.74} & -2.56\\
   \bottomrule
\end{tabular}
\end{adjustbox}
\caption{Rating prediction AUC-ROC. $\Delta$ compares the best [\textbf{Ours}] with the best baseline.}
\label{table:rating_prediction_results}
\vspace{-0.6cm}
\end{table}

\section{Experiments}
We evaluate the proposed method and compare it with conventional as well as LLM-based recommenders.
We aim to answer the following research questions: \textbf{RQ1.} Can our method introduce knowledge into LLMs from new recommendation domains? \textbf{RQ2.} How does our model perform compared to the conventional as well as LLM-based recommenders in retrieval, ranking, and rating prediction? \textbf{RQ3.} How beneficial are the individual proposed tasks? \textbf{RQ4.} What's the effect of varying the size of the backbone LLM?

\subsection{Experimental Setting}
\label{sec:exp_setting}
\textbf{Datasets.} We experiment on three real-world datasets: Amazon Toys \& Games, Beauty, and Sports \& Outdoors \footnote{\url{https://nijianmo.github.io/amazon/}}. Following \citealt{zhou2020s3, geng2022recommendation}, we keep 5-core data and apply leave-one-out evaluation, \textit{i.e.}, for each user purchase sequence (where the interactions are sorted by timestamp in ascending order), the last, the second to the last, and the prior interactions are used for testing, validation, and training, respectively. We present data statistics in Appendix \ref{sec:statistics}. 
\vspace{0.1\baselineskip}

\noindent\textbf{Recommendation Tasks.} We evaluate on three established recommendation tasks: \textbf{retrieval}, which retrieves the ground truth item that a user interacted with from the entire dataset; \textbf{ranking}, which chooses the ground truth item that a user interacted with from a candidate pool of size 100 (1 positive item and 99 negative items sampled based on popularity); \textbf{rating prediction}, which classifies an interaction as either "like" or "dislike" (interactions with ratings > 3 are considered as "like"). We leave the exploration and evaluation of novel recommendation tasks (\textit{e.g.}, explanation generation) to the future, due to a lack of ground-truth data.
\vspace{0.1\baselineskip}

\noindent\textbf{Evaluation Metrics.} For retrieval and ranking, we report top-$k$ Hit Ratio (HR@$k$) and Normalized Discounted Cumulative Gain (NDCG@$k$), where $k$ is set to 5/10 and 1/5/10, respectively. For rating prediction, we report Area Under the Receiver Operating Characteristic Curve (AUC-ROC).
\vspace{0.1\baselineskip}

\noindent\textbf{Models.} We compare to non LLM-based recommenders. For retrieval, we consider sequential recommenders including \textbf{Caser} \cite{tang2018personalized}, which leverages CNNs, \textbf{HGN} \cite{ma2019hierarchical}, which adopts hierarchical gating networks, \textbf{GRU4Rec} \cite{hidasi2015session}, which leverages GRUs \cite{cho2014learning}, \textbf{BERT4Rec} \cite{sun2019bert4rec}, \textbf{FDSA} \cite{zhang2019feature}, \textbf{SASRec} \cite{kang2018self}, $\textbf{S}^{3}\textbf{-Rec}$ \cite{zhou2020s3}, and \textbf{TIGER} \cite{rajput2023recommender}, which leverage self-attention, with TIGER being the current SOTA. For ranking, we consider \textbf{BPR-MF} \cite{rendle2012bpr}, \textbf{BPR-MLP} \cite{cheng2016wide}, and \textbf{SimpleX} \cite{mao2021simplex}, which are collaborative filtering-based method. For rating prediction, we consider \textbf{History}, a naive method that always predicts based on how likely a user likes the training items they purchased, \textbf{DMF} \cite{xue2017deep}, a neural matrix factorization model, and \textbf{Wide\&Deep} \cite{cheng2016wide}, a context-aware method.
Beside, we also consider LLM-based methods including \textbf{P5} \cite{geng2022recommendation}, which fine-tunes T5 \cite{raffel2020exploring} with multi-task recommendation prompts, \textbf{P5-XL}, which fine-tunes FLAN-T5-XL with P5 prompts, \textbf{FLAN-T5-Base/XL} \cite{wei2021finetuned}, which make zero-shot predictions with FLAN-T5-Base or FLAN-T5-XL. We query them with our proposed recommendation-task data samples generated from the test set \footnote{We acknowledge that our retrieval and ranking data samples (examples are shown in Figure \ref{fig:overview} and Appendix \ref{sec:appendix_samples}) utilize item IDs for matching prediction results, whereas the FLAN-T5-Base/XL models, when queried in the zero-shot setting, do not inherently predict item IDs. Addressing this discrepancy, text-based methods could be employed to extract item titles, descriptions, etc., from the FLAN-T5-Base/XL predictions to enhance their performance. However, employing such approaches requires an additional model for text matching, which falls beyond the scope of this work}.
\textbf{ReAT/ RaAT/ RpAT}, which fine-tune FLAN-T5-XL with our proposed retrieval (Re), ranking (Ra), or rating prediction (Rp) task data samples along with the auxiliary-task (AT) data samples \footnote{BPR data samples are used only by RaAT as we observe that they help ranking but not retrieval and rating prediction. MIM/ MLM data samples are used by ReAT, RaAT, and RpAT.}, \textbf{unified training (UT)}, which fine-tunes FLAN-T5-XL with a combination of our proposed Re, Ra, Rp data samples, \textbf{unified training w/ auxiliary tasks (UT+AT)}, which fine-tunes FLAN-T5-XL with a combination of our proposed Re, Ra, Rp, MIM, MLM data samples.
\vspace{0.4\baselineskip}

\noindent\textbf{Implementation Details.} We adopt the 3B FLAN-T5-XL \cite{wei2021finetuned} as the backbone. We also use the 223M FLAN-T5-Base for the ablation studies in Section \ref{sec:ablation}. Meanwhile, it's crucial to emphasize that the proposed method is not tied to a specific backbone architecture and is easily adaptable to other LLMs, such as LLaMA \cite{touvron2023llama}.
We set the sliding window size $w$ to 20. For the BPR data samples, we sample the negative items based on popularity. For the ranking and BPR data samples, the position of the positive item in the candidate pool is always determined randomly. For the MIM and MLM data samples, we adopt a masking ratio of 20\%. To fully fine-tune the LLM backbone, we apply dynamic sampling for the BPR and MIM/MLM data samples (we present details about the dynamic sampling and the statistics of our data samples in Appendix \ref{sec:appendix_samples}). To reduce cost, we validate on 3,000 users. Meanwhile, testing is performed on all users.
We fine-tune FLAN-T5-XL and FLAN-T5-Base for $70,000$ and $10,000$ steps, with batch sizes 16 and 64, respectively.
We set the learning rate to 0.001 and warm-up steps to 1,000.
During prediction, we set the width of the beam search for retrieval and ranking to 20.
For unified models, \textit{i.e.}, UT and UT+AT, model selections are based on retrieval validation performance. We present the detailed settings of P5-XL experiments in Appendix \ref{sec:appendix_p5}. We cite the results of some baseline models from \citealt{zhou2020s3, geng2022recommendation, rajput2023recommender}. We implement DMF and Wide\&Deep with RecBole~\footnote{\url{https://recbole.io}}. We adopt the default configurations, except the data split, mapping (ratings to "like"s or "dislike"s), and metric are adjusted to follow our experiment settings as reported earlier. The pseudo code for generating our proposed data samples can be found in Appendix~\ref{sec:appendix_samples}.

\subsection{Overall Performance (RQ1 \& RQ2)}
Tables \ref{table:retrieval_results}, \ref{table:ranking_results}, and \ref{table:rating_prediction_results} show the results of retrieval, ranking, and rating prediction, respectively. FLAN-T5-Base/XL exhibit suboptimal performance on retrieval and ranking. For retrieval, they show near zero NDCGs and HRs. For ranking, they are significantly inferior to the conventional baselines. For rating prediction, they perform much higher than random guessing (50.00), outperforming DMF, but still fall behind History and Wide\&Deep. 
This shows that FLAN-T5 models lack recommendation knowledge, which is unsurprising considering they were not trained on recommendation tasks during pre-training or instruction-tuning and are evaluated in a zero-shot setting.
Moreover, we find that our proposed method effectively aligns LLMs with new recommendation domains (RQ1). In particular, by fine-tuning FLAN-T5-XL with our proposed data samples, our models significantly outperform FLAN-T5-XL on all three tasks across the datasets.

When compared to the baselines, our models show remarkable performance, especially on retrieval (RQ2). For retrieval, our ReAT outperforms TIGER, the current SOTA, by large margins across datasets and metrics.
Additionally, it is essential to highlight that our method possesses natural language reasoning potentials of LLMs, which are absent in TIGER.
For ranking, our RaAT greatly outperforms SimpleX, the best baseline, on Toys \& Games.
On Beauty, RaAT performs on par with SimpleX. On Sports \& Outdoors, RaAT is inferior to the conventional recommenders on metrics such as NDCG/HR@10, yet still greatly outperforms the LLM-based baselines. Notably, the @1 performance of RaAT is always much higher than the conventional recommenders.
For rating prediction, our RpAT outperforms Wide\&Deep, the best baseline, on Toys \& Games and Beauty while lags slightly behind it on Sports \& Outdoors. These results verify that our method introduces substantial recommendation domain knowledge into LLMs for outperforming strong baselines.
The relative ineffectiveness of our method on Sports \& Outdoors for the ranking and rating prediction tasks could be due to the nature of the data. Specifically, our model, as a sequential recommender, relies on the sequential item correlations conveyed by the user sequences. Such signals may be relatively weak in Sports \& Outdoors (\textit{e.g.}, the average sequence length of Sports \& Outdoors is $8.32\pm6.07$, whereas that of Beauty and Toys \& Games are $8.88\pm8.16$ and $8.63\pm8.51$, respectively, suggesting that Sports \& Outdoors sequences are shorter and less diverse), causing our method to perform suboptimally. The best baselines, on the other hand, do not rely on such information. \textit{E.g.}, SimpleX is based on collaborative filtering and Wide\&Deep is a context-based model. Therefore, their performances are not impacted.

Moreover, our UT greatly outperforms P5 and P5-XL across datasets and metrics. This shows that our proposed recommendation task prompts better preserve item correlations as compared to the P5 ones. Specifically, we enhance user sequence modeling by introducing helpful details such as item titles while excluding less informative details such as user IDs and explanation data. Additional results of P5-XL as well as a comparison between P5-XL and P5 can be found in Appendix \ref{sec:appendix_p5}.

We also compare our UT+AT model with our task-specific models, \textit{i.e.}, ReAT/ RaAT/ RpAT. We show that our method allows fine-tuning a unified model that addresses all recommendation tasks without sacrificing per-task performance by much. For retrieval, UT+AT is slightly worse than ReAT but still outperforms all baselines, except that UT+AT performs comparably with TIGER on Sports \& Outdoors. For ranking, UT+AT performs on par with or slightly better than our task-specific RaAT model. For rating prediction, UT+AT is slightly worse than RpAT.

\begin{table}[t]
\centering
\begin{adjustbox}{width=0.85\linewidth}
\begin{tabular}{clcccc}
    \toprule
    \# & Methods & \makecell{NDCG\\@5} & \makecell{NDCG\\@10} & \makecell{HR\\@5} & \makecell{HR\\@10} \\
   \midrule
   1 & TIGER & \underline{0.0371} & \underline{0.0432} & \underline{0.0521} & \underline{0.0712} \\
   \midrule
   2 & FLAN-T5-XL & $2\mathrm{e}{-5}$ & $2\mathrm{e}{-5}$ &  $5\mathrm{e}{-5}$ & $5\mathrm{e}{-5}$ \\
   3 & 2+retrieval & 0.0182 & 0.0219 & 0.0273 & 0.0388 \\
   4 & 3+MLM & 0.0306 & 0.0369 & 0.0443 & 0.0641 \\
   5 & 4+MIM & \textbf{0.0390} & \textbf{0.0461} & \textbf{0.0558} & \textbf{0.0776} \\
   \midrule
    6 & FLAN-T5-Base & 0.0000 & $2\mathrm{e}{-5}$ & 0.0000 & $5\mathrm{e}{-5}$ \\
    7 & 6+retrieval & 0.0149 & 0.0183 & 0.0219 & 0.0325 \\
    8 & 7+MLM & 0.0219 & 0.0271 & 0.0334 & 0.0495 \\
    9 & 8+MIM & 0.0242 & 0.0304 & 0.0376 & 0.0566 \\
    \bottomrule
    \end{tabular}
    \end{adjustbox}
\caption{Retrieval ablation study on Toys \& Games. Rows 1, 2, 5 (equivalent to ReAT), and 6 are copied from Table~\ref{table:retrieval_results}.}
\label{table:ablation_retrieval}
\end{table}

\begin{table}[t]
\centering
\begin{adjustbox}{width=0.95\linewidth}
\begin{tabular}{clccccc}
    \toprule
    \# & Methods & \makecell{NDCG\\@5} & \makecell{NDCG\\@10} & \makecell{HR\\@1} & \makecell{HR\\@5} & \makecell{HR\\@10} \\
   \midrule
   1 & SimpleX & 0.1244 & 0.1469 & 0.0268 & 0.1958 & 0.2662 \\
   \midrule
   2 & FLAN-T5-XL & 0.0160 & 0.0312 & 0.0026 & 0.0315 & 0.0793 \\
   3 & 2+ranking & 0.1520 & 0.1864 & 0.0807 & 0.2218 & 0.3284 \\
   4 & 3+MLM & 0.1580 & 0.1912 & 0.0854 & 0.2303 & \underline{0.3333} \\
   5 & 4+MIM & \underline{0.1677} & \underline{0.1976} & \underline{0.0938} & \underline{0.2391} & 0.3317 \\
   6 & 5+BPR & \textbf{0.1714} & \textbf{0.2034} & \textbf{0.0956} & \textbf{0.2464} & \textbf{0.3453} \\
   \midrule
    7 & FLAN-T5-Base & 0.0107 & 0.0127 & 0.0057 & 0.0156 & 0.0217 \\
    8 & 7+ranking & 0.1349 & 0.1654 & 0.0720 & 0.1957 & 0.2901 \\
    9 & 8+MLM & 0.1481 & 0.1782 & 0.0820 & 0.2119 & 0.3051 \\
    10 & 9+MIM & 0.1489 & 0.1811 & 0.0817 & 0.2141 & 0.3136 \\
    11 & 10+BPR & 0.1534 & 0.1844 & 0.0844 & 0.2196 & 0.3153 \\
    \bottomrule
    \end{tabular}
    \end{adjustbox}
\caption{Ranking ablation study on Toys \& Games. Rows 1, 2, 6 (equivalent to RaAT), and 7 are copied from Table \ref{table:ranking_results}.}
\label{table:ablation_ranking}
\vspace{-0.5cm}
\end{table}


\begin{table}
\centering
\begin{adjustbox}{width=0.45\linewidth}
\begin{tabular}{clc}
\toprule
\# & Methods & AUC-ROC \\
\bottomrule
1 & Wide\&Deep & 70.93 \\
\bottomrule
2 & FLAN-T5-XL & 55.23 \\
3 & 2+rating-prediction & 70.38 \\
4 & 3+MLM &  \underline{71.08} \\
5 & 4+MIM & \textbf{71.16} \\
\bottomrule
\end{tabular}
\end{adjustbox}
\hfill
\centering
\begin{adjustbox}{width=0.5\linewidth}
\begin{tabular}{clc}
\toprule
\# & Methods & AUC-ROC \\
\midrule
6 & FLAN-T5-Base & 57.85 \\
7 & 6+rating-prediction & 69.17 \\
8 & 7+MLM & 67.31 \\
9 & 8+MIM & 68.24 \\
\bottomrule
\end{tabular}
\end{adjustbox}
\caption{Rating-prediction ablation study on Toys \& Games. Rows 1, 2, 5 (equivalent to RpAT), and 6 are copied from Table \ref{table:rating_prediction_results}.}
\label{table:ablation_rating_prediction}
\vspace{-0.5cm}
\end{table}
\subsection{Ablation Studies (RQ3 \& RQ4)}
\label{sec:ablation}
Tables \ref{table:ablation_retrieval}, \ref{table:ablation_ranking}, and \ref{table:ablation_rating_prediction} show ablation studies on Toys \& Games for retrieval, ranking, and rating prediction, respectively. 
We observe that all the proposed tasks are beneficial (RQ3). In Table~\ref{table:ablation_retrieval} rows 2-5, successively adding our proposed retrieval, MLM, and MIM data samples into the fine-tuning data increases the retrieval performance. All three tasks are essential. \textit{E.g.}, row 4, which fine-tunes FLAN-T5-XL using retrieval and MLM data samples performs on par with S\textsuperscript{3}-Rec and worse than TIGER (row 1, the current SOTA). Further adding MIM data samples (row 5) surpasses TIGER. This shows that the item-level and token-level item correlations introduced by MIM and MLM are essential and complement each other. Similarly, in Table~\ref{table:ablation_ranking} rows 2-6, the ranking performance improves as we incorporate our proposed ranking, MLM, MIM, and BPR data samples into fine tuning. Among these data samples, ranking task data samples are the most helpful. BPR data samples, which contrast the positive items with the negative ones, provide the least assistance.
For rating predictions, as shown in Table~\ref{table:ablation_rating_prediction} rows 2-5, our proposed rating prediction data samples greatly increase the performance. MLM and MIM do help, but only marginally. 

We also find that our proposed method is effective regardless of the size of the backbone model (RQ4). In Tables~\ref{table:ablation_retrieval}, \ref{table:ablation_ranking}, and \ref{table:ablation_rating_prediction}, we apply our method on FLAN-T5-Base and observe significant performance increases on all three recommendation tasks. In terms of overall performance, our best retrieval model with FLAN-T5-Base (Table~\ref{table:ablation_retrieval} row 9) falls behind TIGER but still outperforms all baselines except TIGER, S\textsuperscript{3}-Rec, and SASRec. In Table~\ref{table:ablation_ranking}, our best ranking model with FLAN-T5-Base (row 11) outperforms SimpleX by large margins, though falls behind our best ranking model with FLAN-T5-XL (row 6). In Table~\ref{table:ablation_rating_prediction}, our best rating prediction model with FLAN-T5-Base (row 7) is slightly inferior to the best model with FLAN-T5-XL (row 5) and Wide\&Deep. The effectiveness of the individual tasks remains roughly consistent with the previous results with FLAN-T5-XL (except that MLM does not help rating prediction). \textit{E.g.}, in Table~\ref{table:ablation_ranking} rows 7-11, our ranking task, MLM, MIM, and BPR data samples all contribute to the ranking performance, with the ranking task data samples being the most beneficial and BPR the least beneficial.

\section{Conclusion}
We propose to align LLMs with the recommendation domain by fine-tuning with data samples that encode recommendation knowledge. We propose auxiliary-task data samples that encode item correlations contained in users' preferences.
We further design recommendation-task data samples that are more informative than ones in existing studies. Experiments on retrieval, ranking, and rating prediction show that our method effectively introduces recommendation knowledge into FLAN-T5-Base/XL from three domains. Our method greatly outperforms both conventional and LLM-based baselines in retrieval, achieving the new SOTA.


\section{Limitations}
Our proposed method utilizes LLMs as the backbones. The substantial parameter size of the LLMs results in increased computational resource consumption and extended training and inference times compared to conventional recommenders.
Nevertheless, adopting LLM backbones is beneficial due to their significant potential. In addition to the exceptional performance demonstrated in this study, we anticipate that future research will continue to augment existing recommendation tasks and address novel recommendation scenarios by leveraging the diverse capabilities of LLM backbones.

\bibliography{acl_latex}

\clearpage 
\appendix

\begin{table}[t]
\scriptsize
\centering
\begin{adjustbox}{width=\linewidth}
\begin{tabular}{l|cccc}
\hline
Dataset & \# Users & \# Items & \# Interactions & Sparsity (\%) \\
\hline
Toys \& Games & 19,412 & 11,924 & 167,597 & 99.93 \\
Beauty & 22,363 & 12,101 & 198,502 & 99.93 \\
Sports \& Outdoors & 35,598 & 18,357 & 296,337 & 99.95 \\
\hline
  \end{tabular}
\end{adjustbox}
\caption{Statistics of the datasets.}
\label{table:statistics}
\end{table}

\begin{table*}[t]
\centering
\begin{adjustbox}{width=\linewidth}
\begin{tabular}{lccccccccccccccc}
    \toprule
    \multirow{2.5}{*}{Methods} &
    \multicolumn{5}{c}{\textbf{Toys \& Games}} & \multicolumn{5}{c}{\textbf{Beauty}} & \multicolumn{5}{c}{\textbf{Sports \& Outdoors}} \\
   \cmidrule(lr){2-6}\cmidrule(lr){7-11}\cmidrule(lr){12-16}
   & \shortstack{NDCG\\@5}  & \shortstack{NDCG\\@10} & \shortstack{HR\\@1} & \shortstack{HR\\@5}  & \shortstack{HR\\@10}
   & \shortstack{NDCG\\@5}  & \shortstack{NDCG\\@10} & \shortstack{HR\\@1} & \shortstack{HR\\@5}  & \shortstack{HR\\@10}
   & \shortstack{NDCG\\@5}  & \shortstack{NDCG\\@10} & \shortstack{HR\\@1} & \shortstack{HR\\@5}  & \shortstack{HR\\@10} \\
   \cmidrule{1-16}
   P5-XL & \underline{0.0290} & \underline{0.0444} & \underline{0.0097} & \underline{0.0494} & \underline{0.0977} & \underline{0.0298} & \underline{0.0456} & \underline{0.0110} & \underline{0.0498} & \underline{0.0992} & \underline{0.0286} & \underline{0.0436} & \underline{0.0097} & \underline{0.0486} & \underline{0.0957} \\
   P5-XL (5-5) & 0.0274 & 0.0428 & 0.0089 & 0.0467 & 0.0948 & 0.0289 & 0.0443 & 0.0093 & 0.0497 & 0.0982 & 0.0275 & 0.0426 & 0.0091 & 0.0470 & 0.0943 \\
   \midrule
   UT [\textbf{Ours}] & \textbf{0.1536} & \textbf{0.1867} & \textbf{0.0831} & \textbf{0.2233} & \textbf{0.3259} & \textbf{0.1236} & \textbf{0.1537} & \textbf{0.0609} & \textbf{0.1863} & \textbf{0.2798} & \textbf{0.0867} & \textbf{0.1137} & \textbf{0.0381} & \textbf{0.1362} & \textbf{0.2202} \\
   \bottomrule
   \end{tabular}
   \end{adjustbox}
\caption{Additional P5-XL Ranking results. Rows 1 and 3 are copied from Table \ref{table:ranking_results}.}
\label{table:additional_ranking_results}
\end{table*}

\begin{table}[t]
\centering
\begin{adjustbox}{width=0.85\linewidth}
\begin{tabular}{lcccc}
    \toprule
    Methods & \makecell{NDCG\\@5} & \makecell{NDCG\\@10} & \makecell{HR\\@5} & \makecell{HR\\@10} \\
   \midrule
   UT [\textbf{Ours}] & \textbf{0.0079} & \textbf{0.0101} & \underline{0.0118} & \textbf{0.0187} \\
   UT+IE [\textbf{Ours}] & \underline{0.0076} & \underline{0.0097} & \textbf{0.0121} & \underline{0.0185} \\
    \bottomrule
    \end{tabular}
    \end{adjustbox}
\caption{Retrieval results on Sports \& Outdoors with (UT+IE) or without (UT) IE data samples. Row 1 is copied from Table~\ref{table:retrieval_results}.}
\label{table:IE_retrieval}
\end{table}

\begin{table*}[t]
\centering
\begin{adjustbox}{width=0.7\linewidth}
\begin{tabular}{lccccccccc}
    \toprule
    \multirow{2.5}{*}{Task} &
    \multicolumn{3}{c}{\textbf{Toys \& Games}} & \multicolumn{3}{c}{\textbf{Beauty}} & \multicolumn{3}{c}{\textbf{Sports \& Outdoors}} \\
   \cmidrule(lr){2-4}\cmidrule(lr){5-7}\cmidrule(lr){8-10}
   & \shortstack{\# Train}  & \shortstack{\# Valid} & \shortstack{\# Test}
   & \shortstack{\# Train}  & \shortstack{\# Valid} & \shortstack{\# Test} 
   & \shortstack{\# Train}  & \shortstack{\# Valid} & \shortstack{\# Test} \\
   \cmidrule{1-10}

   Retrieval & 30,761 & 3,000 & 19,412 & 36,582 & 3,000 & 22,363 & 47,320 & 3,000 & 35,598 \\
   Ranking & 30,761 & 3,000 & 19,412 & 36,582 & 3,000 & 22,363 & 47,320 & 3,000 & 35,598 \\
   Rating prediction & 30,761 & 3,000 & 19,412 & 36,582 & 3,000 & 22,363 & 47,320 & 3,000 & 35,598 \\
   
   \midrule
   MIM & DS & 0 & 0 & DS & 0 & 0 & DS & 0 & 0\\
   MLM & DS & 0 & 0 & DS & 0 & 0 & DS & 0 & 0\\
   BPR & DS & 0 & 0 & DS & 0 & 0 & DS & 0 & 0\\
   
   \bottomrule
   \end{tabular}
   \end{adjustbox}
\caption{Statistics of our proposed data samples. DS stands for dynamic sampling.}
\label{table:statistics_proposed_task}
\end{table*}

\section{P5-XL Experimental Setting and Additional Results}
\label{sec:appendix_p5}
\subsection{Experimental Setting}
We generate P5 prompts using the source code provided by the P5 authors \footnote{\url{https://github.com/jeykigung/P5}}. However, for a fair comparison, we update the data pre-processing to be consistent with our method and the other baselines. Specifically, we apply random instead of sequential indexing when mapping the item IDs. As pointed out by \citealt{rajput2023recommender}, the sequential indexing of items (\textit{e.g.}, the purchase sequence of the first user in Toys \& Games is mapped into `1, 2, 3, 4, 5, 6, 7') in the original P5 pre-processing leads to data leakage (\textit{e.g.}, given the train items, \textit{i.e.}, `1, 2, 3, 4, 5, 6', the LLM can easily infer the test item, \textit{i.e.}, `7'). Therefore, we adopt random mapping (\textit{i.e.}, consecutive or similar-looking IDs may not imply any connection), which is consistent with our method. In addition, the original P5 pre-processing adopts leave-one-out split for retrieval and ranking, while splitting the dataset by 0.8:0.1:0.1 for the training, validation, and testing of rating prediction. This could result in data leakage, as the test interactions of one task might be included in the training set of another task.
We instead adopt leave-one-out data split for all three recommendation tasks, which is consistent with our proposed method as well as the other baselines. 

For a fair comparison, We apply the same backbone (FLAN-T5-XL), fine-tuning steps (70,000), batch size (16), and learning rate (0.001) as adopted by our proposed method. Following the original P5 code, we fine-tune a unified model with prompts of their proposed five task families (rating, sequential recommendation, explanation, review, and direct recommendation. The sequential recommendation and direct recommendation families are weighted 5 times higher than the rest families). In Tables \ref{table:retrieval_results}, \ref{table:ranking_results}, and \ref{table:rating_prediction_results}, we adopt prompt templates 2-1, 2-7, and 1-4 for evaluating the retrieval, ranking, and rating prediction performance of the P5-XL model, as these templates better suit the forms of the recommendation tasks (introduced in the second subsection of Section \ref{sec:exp_setting}) than the other templates. 

\subsection{P5-XL vs. P5}
Please note that the retrieval results of P5 in Table \ref{table:retrieval_results} are cited from \citealt{rajput2023recommender} rather than the original P5 paper \cite{geng2022recommendation}. This is because the original P5 experiments cannot be reproduced upon fixing the information leakage issues as discussed in the previous section. Meanwhile, \citealt{rajput2023recommender} does not report the ranking and rating prediction performances of P5. To fully evaluate P5, we train a P5-XL model following the experimental setting as detailed in the previous section, and report its performance on all three tasks in Tables \ref{table:retrieval_results} to \ref{table:rating_prediction_results}.

P5-XL performs worse than P5 in Table \ref{table:retrieval_results}, which is likely owing to the differences in their training data. Specifically, P5 was only trained on retrieval prompts (as indicated in Appendix D of \citealt{rajput2023recommender}). While following the original P5 paper, P5-XL is trained on all five task families of P5 prompts, including explanation generation and review summarization tasks. We hypothesize that these additional data samples are very different from the evaluated tasks (retrieval, ranking and rating prediction), causing negative transfer to the evaluated tasks.

\subsection{Additional Results}
In Table \ref{table:additional_ranking_results}, we report the ranking results of P5-XL evaluated with prompt template 5-5. We can tell that P5-XL (5-5) slightly fall behind P5-XL. Our proposed UT greatly outperforms both P5-XL and P5-XL (5-5), which again verifies that our proposed recommendation task prompts are more informative than the P5 ones.

\section{Dataset Statistics}
\label{sec:statistics}
Table \ref{table:statistics} presents the statistics of the Amazon datasets, \textit{i.e.}, Toys \& Games, Beauty, and Sports \& Outdoors, that we used to evaluate our proposed method as well as all the baselines.

\section{Pseudo Code, Statistics, and Examples of the Proposed Data Samples}
\label{sec:appendix_samples}
\subsection{Pseudo Code for Data Sample Generation}
Algorithm \ref{algorithm:data_sample_generation} presents the pseudo code for generating our proposed recommendation-task and auxiliary-task data samples.
\subsection{Statistics of the Data Samples}
Table \ref{table:statistics_proposed_task} presents the statistics of our proposed recommendation-task and auxiliary-task data samples. Consider the recommendation-task data samples, the training data samples are generated by swiping a sliding window of size $w=20$ over the training split of the user sequence. The validation data samples consider only 3,000 users for each dataset for cost-efficient validation. We test on all users, therefore the counts of the testing data samples equal to the total number of users in the datasets. The auxiliary-task data samples, on the other hand, are generated using only the training splits. Notably, during training, we apply dynamic sampling that decide the negative items in the BPR data samples as well as the masked items/tokens in the MIM/MLM data samples on the fly. Such dynamic sampling helps to fully fine-tune the LLM backbones.

\subsection{Examples of the Data Samples}
In Table \ref{table:samples}, we present examples of our proposed data samples. These data samples are generated with the training data split of an Amazon - Toys \& Games user whose ID is `A12HF3UBDV34RR'. Note that to fully fine-tune the LLM backbone, we apply dynamic sampling for the BPR and MIM/MLM data samples and decide the negative items and masked items/tokens on the fly. Here, we only present the BPR, MIM, and MLM data samples resulted from a single sampling.

\begin{table*}[t]
\scriptsize
\centering
\begin{adjustbox}{width=\linewidth}
\begin{tabular}{c|p{15cm}}
    \toprule
    Task & Data sample \\
   \midrule
   Retrieval &  \textbf{Input}:  A user has purchased the following Amazon products (arranged in chronological order, from earliest to most recent): Item ID: I9762, Title: Winstonia's 8 Wheels Combo Set Nail Art Polymer Slices Fimo Decal Pieces Accessories - Butterflies, Bows, Animals, Fruit, Flowers, Dragonflies, Cupcakes, Hearts; Item ID: I8123, Title: MASH Rhinestones 2400 Piece 12 Color Nail Art Nailart Manicure Wheels; Item ID: I158, Title: Aveeno Clear Complexion Daily Moisturizer, 4 Ounce; Item ID: I5324, Title: Bdellium Tools Professional Antibacterial Makeup Brush Studio Line - Precision Kabuki Airbrushed Effect 957; Item ID: I7522, Title: Bdellium Tools Professional Makeup Brush Green Bambu Series Smoky Eyes 5pc. Brush Set; Item ID: I7647, Title: real Techniques Stippling Brush; Item ID: I7811, Title: Maybelline New York Color Sensational High Shine Lipcolor, Coral Lustre 840, 0.12 Ounce; Item ID: I9440, Title: Bed Head BH313 Orange Crush 1-inch Styler; Item ID: I5046, Title: Herstyler Baby Curl Curling Iron, Purple; What would the user buy next?

    \textbf{Output}: I3977\\
   Ranking & \textbf{Input}:  A user has purchased the following Amazon products (arranged in chronological order, from earliest to most recent): Item ID: I9762, Title: Winstonia's 8 Wheels Combo Set Nail Art Polymer Slices Fimo Decal Pieces Accessories - Butterflies, Bows, Animals, Fruit, Flowers, Dragonflies, Cupcakes, Hearts; Item ID: I8123, Title: MASH Rhinestones 2400 Piece 12 Color Nail Art Nailart Manicure Wheels; Item ID: I158, Title: Aveeno Clear Complexion Daily Moisturizer, 4 Ounce; Item ID: I5324, Title: Bdellium Tools Professional Antibacterial Makeup Brush Studio Line - Precision Kabuki Airbrushed Effect 957; Item ID: I7522, Title: Bdellium Tools Professional Makeup Brush Green Bambu Series Smoky Eyes 5pc. Brush Set; Item ID: I7647, Title: real Techniques Stippling Brush; Item ID: I7811, Title: Maybelline New York Color Sensational High Shine Lipcolor, Coral Lustre 840, 0.12 Ounce; Item ID: I9440, Title: Bed Head BH313 Orange Crush 1-inch Styler; Item ID: I5046, Title: Herstyler Baby Curl Curling Iron, Purple; Which of the following candidate items would you recommend the user to buy next? Candidate items are: I10537, I11849, I2647, I10506, I377, I8136, I3598, I2316, I114, I10379, I6767, I2801, I4687, I3446, I7222, I5925, I4608, I2226, I2279, I11708, I4376, I8771, I6502, I8650, I7006, I11350, I6716, I4690, I11303, I3446, I8704, I4001, I9816, I1498, I6896, I1598, I7653, I2086, I12019, I3235, I12052, I27, I5786, I9936, I697, I10050, I447, I10898, I2093, I2618, I2044, I2618, I6924, I2769, I8117, I10772, I9252, I4668, I6982, I2234, I9894, I9441, I6514, I5519, I8620, I710, I10212, I8654, I7648, I11054, I1419, I10958, I334, I576, I1537, I8278, I3181, I189, I3510, I7974, I6010, I11187, I6465, I9596, I9356, I311, I2313, I7117, I9249, I643, I6732, I8803, I5499, I2434, I3977, I10691, I10707, I5553, I7999, I8672.
   
\textbf{Output}: I3977\\
   Rating prediction & \textbf{Input}:  A user likes the following Amazon products: Item ID: I7522, Title: Bdellium Tools Professional Makeup Brush Green Bambu Series Smoky Eyes 5pc. Brush Set; Item ID: I7811, Title: Maybelline New York Color Sensational High Shine Lipcolor, Coral Lustre 840, 0.12 Ounce; The user dislikes the following Amazon products: Item ID: I7647, Title: real Techniques Stippling Brush; Item ID: I9440, Title: Bed Head BH313 Orange Crush 1-inch Styler; Item ID: I5046, Title: Herstyler Baby Curl Curling Iron, Purple; Predict whether the user would like the following item. Answer yes or no. Item ID: I3977, Title: L'Oreal Paris HiP Studio Secrets Professional Color Truth Cream Eyeliner, Brown, 0.159 Ounce
   
    \textbf{Output}:  no \\
    \midrule
    MIM & \textbf{Input}: A user has purchased the following Amazon products (arranged in chronological order, from earliest to most recent): Item ID: I9762, Title: Winstonia's 8 Wheels Combo Set Nail Art Polymer Slices Fimo Decal Pieces Accessories - Butterflies, Bows, Animals, Fruit, Flowers, Dragonflies, Cupcakes, Hearts; [masked item]; Item ID: I158, Title: Aveeno Clear Complexion Daily Moisturizer, 4 Ounce; Item ID: I5324, Title: Bdellium Tools Professional Antibacterial Makeup Brush Studio Line - Precision Kabuki Airbrushed Effect 957; Item ID: I7522, Title: Bdellium Tools Professional Makeup Brush Green Bambu Series Smoky Eyes 5pc. Brush Set; [masked item]; Item ID: I7811, Title: Maybelline New York Color Sensational High Shine Lipcolor, Coral Lustre 840, 0.12 Ounce; Item ID: I9440, Title: Bed Head BH313 Orange Crush 1-inch Styler; Item ID: I5046, Title: Herstyler Baby Curl Curling Iron, Purple; Item ID: I3977, Title: L'Oreal Paris HiP Studio Secrets Professional Color Truth Cream Eyeliner, Brown, 0.159 Ounce; What are the masked items, in chronological order?
    
    \textbf{Output}: Item ID: I8123, Title: MASH Rhinestones 2400 Piece 12 Color Nail Art Nailart Manicure Wheels; Item ID: I7647, Title: real Techniques Stippling Brush; \\
    MLM & \textbf{Input}: Item ID: I7811, Title: Maybelline New York Color Sensational High Shine Lipcolor, Coral Lustre 840, 0.12 Ounce; Item ID: I9440, Title: Bed Head BH313 Orange Crush 1-inch Styler;\\
    BPR & \textbf{Input}: A user has purchased the following Amazon products (arranged in chronological order, from earliest to most recent): Item ID: I9762, Title: Winstonia's 8 Wheels Combo Set Nail Art Polymer Slices Fimo Decal Pieces Accessories - Butterflies, Bows, Animals, Fruit, Flowers, Dragonflies, Cupcakes, Hearts; Item ID: I8123, Title: MASH Rhinestones 2400 Piece 12 Color Nail Art Nailart Manicure Wheels; Item ID: I158, Title: Aveeno Clear Complexion Daily Moisturizer, 4 Ounce; Item ID: I5324, Title: Bdellium Tools Professional Antibacterial Makeup Brush Studio Line - Precision Kabuki Airbrushed Effect 957; Item ID: I7522, Title: Bdellium Tools Professional Makeup Brush Green Bambu Series Smoky Eyes 5pc. Brush Set; Item ID: I7647, Title: real Techniques Stippling Brush; Item ID: I7811, Title: Maybelline New York Color Sensational High Shine Lipcolor, Coral Lustre 840, 0.12 Ounce; Item ID: I9440, Title: Bed Head BH313 Orange Crush 1-inch Styler; Item ID: I5046, Title: Herstyler Baby Curl Curling Iron, Purple; Which of the following two items would the user buy next? Item ID: I4168, Title: Sulfur Soap with Lanolin; Item ID: I3977, Title: L'Oreal Paris HiP Studio Secrets Professional Color Truth Cream Eyeliner, Brown, 0.159 Ounce; 
    
    \textbf{Output}: Item ID: I3977, Title: L'Oreal Paris HiP Studio Secrets Professional Color Truth Cream Eyeliner, Brown, 0.159 Ounce;\\
    
   \bottomrule
   \end{tabular}
   \end{adjustbox}
\caption{Examples of our proposed data samples.}
\label{table:samples}
\end{table*}

\section{Mimicking Item Embedding}
\label{sec:appendix_IE}

\begin{figure}[t]
\centering
\includegraphics[width = 7.7cm]{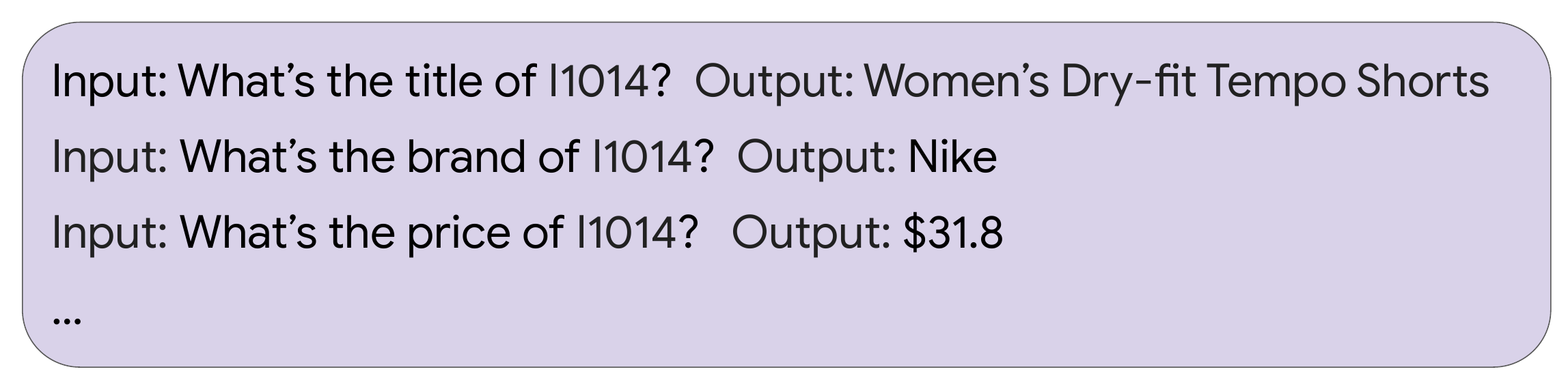}
\caption{\label{fig:IE} Item embedding (IE) data samples.}
\end{figure}

\begin{algorithm}
\caption{Generate Data Samples}\label{algorithm:data_sample_generation}
\KwIn{Raw interactions, data sample templates for recommendation and auxiliary tasks, data\_split $\in$ \{\textit{Train}, \textit{Valid}, \textit{Test}\}, window size $w$, candidate pool size $c$}
\KwOut{Data samples $\mathcal{D}$}
$\mathcal{I} \gets$ a set of unique items (shuffled and mapped to short IDs)

$\mathcal{S} \gets$ a list of chronologically ordered user purchase sequences

$\mathcal{D} \gets \{\}$

\For{$s \in \mathcal{S}$}{
    \If{data\_split = \textit{Train}}{
        $s_{sub} \gets$ all subsequences of the training split of $s$, each is of length up to $w$
    }
    \If{data\_split = \textit{Valid}}{
        $s_{sub} \gets$ a subsequence of $s$ that ends with the validation item, proceeding items beyond $w$ are truncated
    }
    \If{data\_split = \textit{Test}}{
        $s_{sub} \gets$ a subsequence of $s$ that ends with the test item, proceeding items beyond $w$ are truncated
    }
    \For{$ss \in s_{sub}$}{
        \For{task $\in$ \{\textit{Retrieval}, \textit{Ranking}, \textit{Rating prediction}\}}{
            \If{task = \textit{Ranking}}{
                $neg \gets$ sample $c-1$ negative items from $\mathcal{I}\backslash s$
            }
            Generate a data sample $d$ with $ss$, task template, and $neg$ (for \textit{Ranking} only)

            Add $d$ to $\mathcal{D}$
        }
        \If{data\_split = \textit{Train}}{
            \For{task $\in$ \{\textit{MIM}, \textit{MLM}, \textit{BPR}\}}{
                \If{task = \textit{BPR}}{
                    $neg \gets$ sample $1$ negative item from $\mathcal{I}\backslash s$
                }
                Generate a data sample $d$ with $ss$, task template, and $neg$ (for \textit{BPR} only)

                Add $d$ to $\mathcal{D}$
            }
        }
    }

}

\Return{$\mathcal{D}$}
\end{algorithm}

Our proposed data samples introduced in the main paper encode item correlations encompassed in users' preferences. We also explore encoding item correlations encompassed in item contents, \textit{i.e.}, categories, descriptions, etc. 

We observe that the conventional context-aware recommenders commonly integrate item contents to help the model better understand the items and achieve enhanced performance. \textit{E.g.}, \citealt{hou2022towards} embed the concatenations of item content fields with BERT \cite{devlin2018bert}. The learned item embeddings, $\mathbf{X} \in \mathbb{R}^{N\times d}$, where $N$ is the number of the items and $d$ is the dimension of the vector space, serve as initial representations of the items.

We mimic this item embedding (IE) process with natural language prompts. As shown in Figure \ref{fig:IE}, by asking questions about the properties of an item in the input and answering them in the output, we can generate item embedding data samples such as `Input: What’s the brand of I1014? Output: Nike'. We repeat such question answering process for various available item content fields, including title, categories, brand, price, attributes, and descriptions. These data samples represent knowledge about the items, but with natural language rather than numerical vectors. We expect that tuning LLMs with IE data samples can help them to comprehend the items in the target recommendation domain and enhance their performance. 

To evaluate the IE data samples, we tune a \textbf{UT+IE} model, which augments the fine-tuning data of our UT model with IE data samples (the rest experimental settings of UT+IE and UT remain the same). We present its retrieval performance on Sports \& Outdoors in Table \ref{table:IE_retrieval}. We observe no noticeable performance increase when incorporate IE data samples. The reason might be, the raw item content fields are noisy. \textit{E.g.}, the description field is long and can contain noise such as hashtags and URLs. It has been shown \cite{cao2023multi} pre-processing the raw fields to extract fine-grained features helps to enhance context-aware recommenders. Inspired by this, in the future, we plan to improve the IE data samples by refining the item content fields.

\end{document}